\documentclass[aps,prb,preprint,superscriptaddress,floatfix]{revtex4}

\usepackage{graphicx}
\usepackage{dcolumn}
\usepackage{bm}
\usepackage{amsfonts}
\usepackage{amsmath, amsthm, amssymb}

\newcommand{\eqz}[1]{Eq.~(\ref{#1})}
\newcommand{\Fig}[1]{Fig.~\ref{#1}}
\newcommand{\Reff}[1]{Ref.~[\onlinecite{#1}]}
\newcommand{\Sch}{Schr\"odinger}
\newcommand{\im}   {\mathrm{i}}         

\begin{document}

\title{Magnetic states in prismatic core multishell nanowires}

\author{Giulio Ferrari}
\email[]{giulio.ferrari@unimore.it}
\affiliation{CNR-INFM Research Center on nanoStructures and bioSystems at Surfaces (S3), Via Campi 213/A, 41100 Modena, Italy}
\affiliation{CNISM Unit{\`a} di Ricerca di Modena}
\author{Guido Goldoni}
\affiliation{CNR-INFM Research Center on nanoStructures and bioSystems at Surfaces (S3), Via Campi 213/A, 41100 Modena, Italy}
\affiliation{Dipartimento di Fisica, University of Modena and Reggio Emilia}
\author{Andrea Bertoni}
\affiliation{CNR-INFM Research Center on nanoStructures and bioSystems at Surfaces (S3), Via Campi 213/A, 41100 Modena, Italy}
\author{Giampaolo Cuoghi}
\affiliation{Dipartimento di Fisica, University of Modena and Reggio Emilia}
\author{Elisa Molinari}
\affiliation{CNR-INFM Research Center on nanoStructures and bioSystems at Surfaces (S3), Via Campi 213/A, 41100 Modena, Italy}
\affiliation{Dipartimento di Fisica, University of Modena and Reggio Emilia}



\date{\today}

\begin{abstract}
We study the electronic states of core multi-shell semiconductor nanowires, including the effect of strong magnetic fields. We show that
the multi-shell overgrowth of a free-standing nanowire, together with the prismatic symmetry of the substrate, may induce quantum confinement of carriers in a set of quasi-1D quantum channels corresponding to the nanowire edges. Localization and inter-channel tunnel coupling are controlled by the curvature at the edges and the diameter of the underlying nanowire. We also show that a magnetic field may induce either Aharonov-Bohm oscillations of the energy levels in the axial configuration, or a dimensional transition of the quantum states from quasi-1D to Landau levels for fields normal to the axis. Explicit predictions are given for nanostructures based on GaAs, InAs, and InGaN with different symmetries.
\end{abstract}


\maketitle

Semiconductor nanowires (NWs) are emerging as new nano-devices with an enormous potential for applications.
By now, several classes of materials can be grown in the form of single-crystal NWs, and key steps towards nano-electronic applications have been achieved: control of the position, monodisperse diameter, and uniformity of the NWs\cite{Morales98,Ohlsson01,Mohan05,Mandl06}; $n$ and $p-$doping, as well as growth of embedded barriers and superlattices along the growth direction;\cite{Duan01,Gudiksen02,Bjork02}
self-catalatic\cite{Novotny05,Mattila06,Morral08A} or catalyst-free\cite{Poole03,Mohan05,Mandl06} growth which avoids deep traps formation of extrinsic metal-assisted growth\cite{Wagner64,Hiruma93,Morales98};
combination of different NW materials and substrates, including Si.\cite{Mattila06}

High quality free-standing NWs may also be used as substrates for epitaxial overgrowth of core shell structures.
Also core multi-shell NWs have recently been demonstrated for GaN/In$_{x}$Ga$_{1-x}$N\cite{Qian04}, GaAs/AlAs\cite{Morral08} and InP/InAs\cite{Mohan06}.
Here, quasi-2D electronic systems can be formed, as in usual planar multi-layered heterostructures such as quantum wells, but wrapped around the NW.
For example, in the InP/InAs structures\cite{Mohan06} electrons and holes are confined, due to the different band offset, in a few-monolayer shell of InAs overgrown on a InP core.
Further developments in the growth and doping\cite{Cornet07} may lead to an entirely new class of electronic devices and functionalities, based on high-mobility, bent two-dimensional electron gases with axial symmetry, confined in the low gap shells of the nanostructure.
These may compete with carbon-nanotubes in the quest for the ultimate integration of nano-devices for solar cells\cite{Tian07,Zhang07}, opto-electronic applications\cite{Qian04,Jabeen08}, sensing, or phase-based electronics.

NWs typically grow with a prismatic shape, having the $n$-fold axial symmetry of the equivalent stable crystallographic directions. In fact, the stability of specific crystallographic surfaces leads to faceting.
Specifically, III-V semiconductor-based NWs (InP, InAs, GaAs)\cite{Mohan05,Mandl06,Morral08A} often grow as hexagonal crystals, exposing the \{110\} vertical facets normal to the [111] plane, but square (InP, InAs)\cite{Seifert04,Pfund06} or triangular (GaN)\cite{Kuykendall03,Qian04} shapes have also been demonstrated.
Therefore, the electronic system confined in overgrown shells has the $n$-fold symmetry of the NW used as a substrate. For NWs with diameters in the range of few tens of nm, this results in a substantial deviation from the cylindrical shape which one would be tempted to use in modeling such systems. Indeed, one should note that a bent quantum well behaves like a weakly confining quantum wire in the region of the bent.\cite{Kapon89,Goldoni96}
Therefore, the edges of a prismatic NW tend to be regions of preferred localization  even with a constant well width, this effect being concurrent or in competition with possible width fluctuations in the region of the bent. As we show below, the nanostructure may behave as a set of $n$ tunnel-coupled quantum channels, depending on structural parameters at the nanoscale, such as the NW diameter and symmetry or the curvature at the edges.

In this Letter we investigate the electronic states of core multi-shell NWs. Taking explicitly into account their prismatic symmetry, we show that quantum confinement occurs in a set of quasi-1D channels located at the edges of the NW.
An axial magnetic field affects the overall phase of these states and induce Aharonov-Bohm (AB) oscillations of the multiplets arising from the prismatic symmetry. A transverse magnetic field, instead, may deplete these channels and induce Landau level formation along the NW facets.
Results are presented for several devices based on different classes of materials and symmetries.

\begin{figure}
\centering
\includegraphics[width=3cm, angle=0]{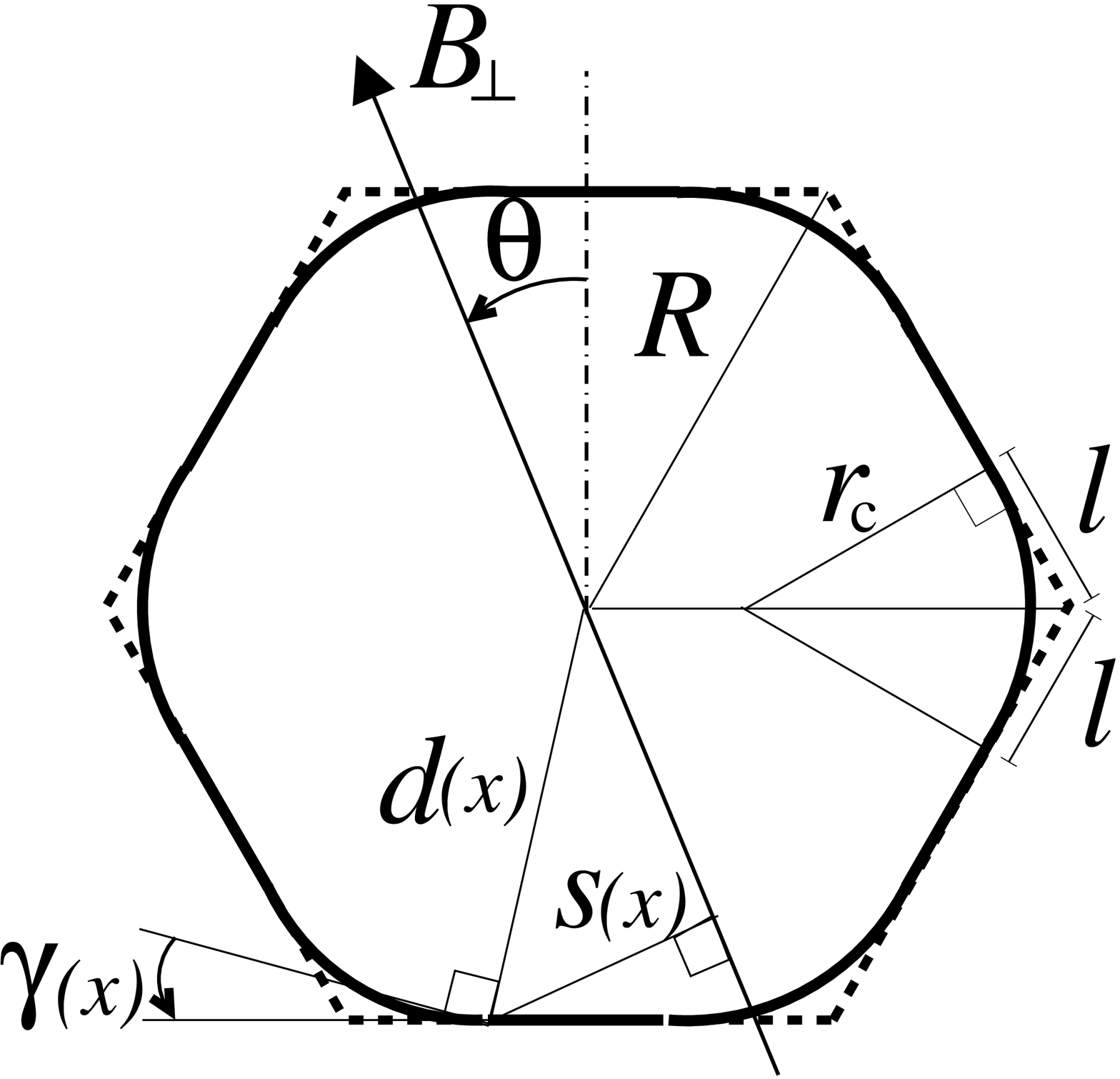}
\includegraphics[width=3.7cm, angle=0]{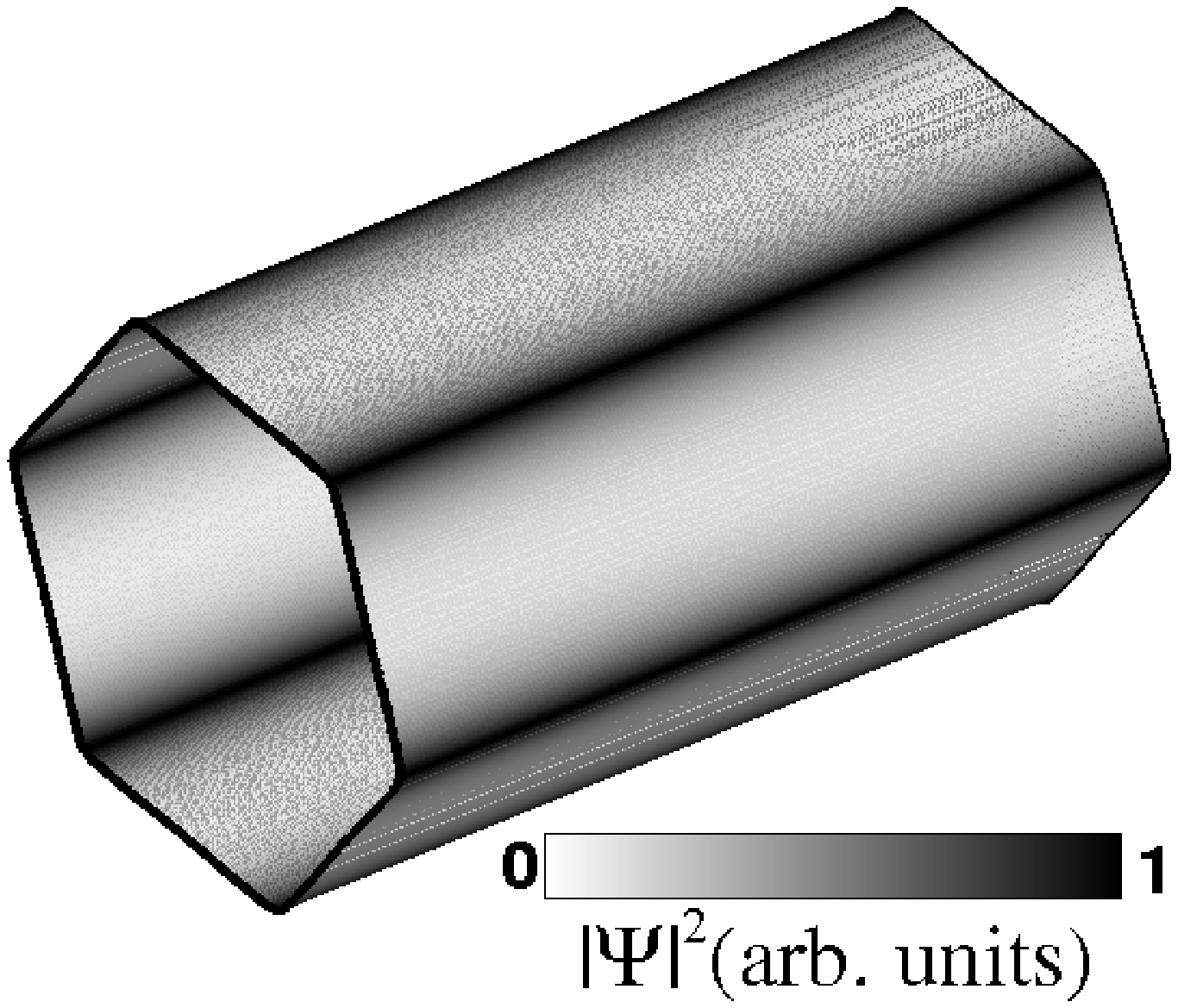}
\caption{\label{fig:esawire}
Left: the cross-section of a prismatic NW is modeled from a regular polygon (dashed line).
The radius of the circumscribed circle is $R$. Each edge is rounded
for a length $l$ and with a constant radius of curvature $r_c = \alpha l$ (see text).
A transverse magnetic field $B_{\bot}$ forms an angle $\theta$ with a line normal to a facet.
The azimuthal position on the rounded polygon (solid line) is given by the $x$ coordinate, at a distance
$d(x)$ from the center. $\gamma(x)$ is the angle between the tangent in $x$ and the line perpendicular to $d(x)$, and $s(x)$ is the distance between $x$ and the direction of the field through the center of the polygon.
Right: charge density of the ground state of a carrier in a core multi-shell NW (gray scale), showing preferred localization along the edges (darker regions). The state shown here is for the GaAs-based nanostructure of \Fig{fig:binding}(b).}
\end{figure}

As we mentioned, a bent quantum well behaves as a 1D electron waveguide in the area of the bent, since there is a larger region to accommodate the wave-function.\cite{Wegscheider93,Rossi97,nota-bent}
This localization effect does not require, and would add to, any fluctuation in the quantum well width (which is likely to happen in real samples\cite{Kapon89}).
Solving the appropriate 3D \Sch\ equation of the nanostructure would be numerically inconvenient in the present large systems, even within the envelope function approximation.
In fact, we are focusing on electronic states which are confined in a layer which is much thinner than the overall large diameter of the NW.
Under these assumptions, the core of the NW acts as substrate for the epitaxial overgrowth, determining the symmetry of the shells.
This allows us to model the bent surface as a strictly 2D curved structure (see \Fig{fig:esawire}).
The confined carriers move under the action of an effective potential which can be derived by a limiting procedure, starting from the full 3D effective-mass \Sch\ equation \cite{Jensen71, daCosta81} and taking the thickness of the bent layer down to zero.
Such curvature-induced effective attractive potential is \cite{Encinosa98, Marchi05}
\begin{equation}\label{eq:potgeo}
V_{\mathrm{bent}}=-{\frac{\hbar^2}{8m^* {r_c}^2}},
\end{equation}
where $m^*$ is the effective mass of the carriers in the shell. The potential depends on the local radius of curvature $r_c$ of the bent.
Note that due to the limiting procedure, electrons sit in a $\delta$-like shell; therefore, the only parameters appearing in the above equation and in the system model presented in \Fig{fig:esawire} are the shell curvature and radius and the effective mass of the carrier corresponding to the shell material.

TEM images\cite{Morral08} show that each edge of the NW actually consists of a region extending over a few nm with a finite curvature. Accordingly, we model the cross-section of a prismatic NW starting from a polygon (see \Fig{fig:esawire}) with a circumscribed circle of radius $R$, and we round off each corner for a length $l$ at each side of the edge. Each rounded edge has a constant curvature $1/r_c$, where $r_c=\alpha l$, with $\alpha = \sqrt{3}$ for the hexagon, $\alpha =1 $ for the square and $\alpha =2/\sqrt{3} $ for the triangle.
A spatial coordinate $x$ is defined which runs on the rounded polygon (solid line in \Fig{fig:esawire}), so that $r_c=r_c(x)$. The prism is therefore modeled
as a $n$-fold succession of planar facets (where $V_{\mathrm{bent}}=0$ since $r_c\rightarrow \infty$) and rounded edges (with a constant attractive potential due to the constant value of $r_c$).
Here we suppose an abrupt change of the curvature between the edges and the prism sides, that is not likely to happen on the atomic scale. Nevertheless, a progressive change between the two curvature would only influence the shape of the well defined by \eqz{eq:potgeo}. The results presented in the following would remain qualitatively unchanged.

\begin{figure}
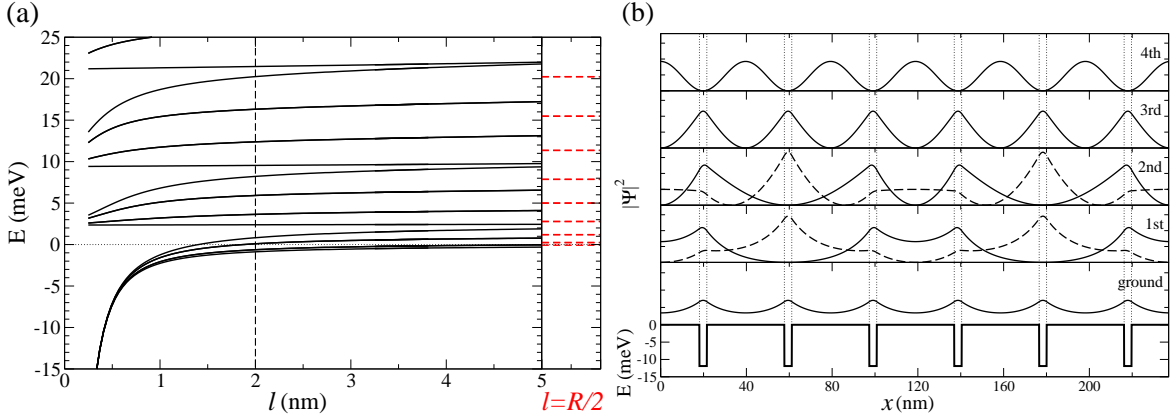

\centering
\includegraphics[height=5.5cm, angle=0]{papernov.eps}
\includegraphics[height=5.5cm, angle=0]{autofunzesa.eps}
\caption{\label{fig:binding} (a) Lowest energy levels in the GaAs layer of an hexagonal core multi-shell NW with $R=40$ nm at zero magnetic field, as a function of $l$, {\em i.e.} of the curvature at the edges.
(b) Confining potential $V_{\mathrm{eff}} (x)$ (bottom panel) and charge densities of the five lowest energy levels $E_{n}$ corresponding to $l=2$ nm  (vertical dashed line of part (a)). Note that the first and second excited levels are doubly degenerate.}
\end{figure}

Since (\ref{eq:potgeo}) does not contain the wave-vector $k_y$ along the axial direction, the energy levels split in a set of parabolic 1D subbands $E_n + \hbar^2 k_y^2/2m^*$, where the $k_y=0$ subband edges $E_n$ are simply determined by a 1D effective mass equation in the step-like potential $V_{\mathrm{eff}}(x) = V_{\mathrm{bent}}(r_c(x))$.
We assume that the curvature of the shell is different form zero only at the rounded edges.
The resulting effective potential has a series of square wells (whose depth is given by \eqz{eq:potgeo}) located at the edges, and separated by regions with $V_{\mathrm{eff}}=0$ corresponding to the prism flat sides.
In \Fig{fig:binding}(a) we show the lowest energy levels $E_n$ of an electron confined in a GaAs\cite{nota-effm} shell overgrown on the surface of an hexagonal NW with $R=40$ nm, as a function of $l$. The energy levels are easily understood as the result of the six-fold attractive potential felt by the carriers, shown in the bottom panel of \Fig{fig:binding}(b) for the realistic value\cite{Morral08} $l=2$ nm. The potential clearly resembles a multiple quantum well $\approx$10 meV deep, with periodic boundary conditions. As shown by the charge density,
\emph{the ground state of the system consists of a set of tunnel-coupled channels located along the edges of the NW, with a confinement energy in the few meV range}.
The probability densities corresponding to the four lowest energy levels show preferred localization at the edges, where the potential is attractive, but with a substantial probability density along the facets of the NW.

For larger $l$ the curvature diminishes, and the quantum wells become wider and shallower, until, for $l=R/2$ (for an hexagonal symmetry), a cylindrical system is obtained, where $V_{\mathrm{bent}}$ just amounts to an addictive constant.
At the opposite limit of small $l$, the wells become thinner and deeper.
The width and depth changes do not compensate, and the confined states depend non trivially on $l$ (or, equivalently, on the curvature).
At the envelope function level used here, an hexagonal NW belongs to the $D_{6h}$ space symmetry group.\cite{Tinkham03}
Accordingly, energy levels group in six-fold bunches, separated by gaps that increase for decreasing $l$, which controls tunneling between the edges.
The ground state is not degenerate, while the first and second excited levels
are doubly degenerate.
The third excited state is again not degenerate and has six nodes in the zero-potential regions.
The topmost graph of \Fig{fig:binding}(b) shows the lowest state of the second bunch, which has maxima located at the centers of the facets of the NW and has nodes in each well.
In a cylindrical system the 3rd and 4th levels would be degenerate \cite{Ferrari08B} but the presence of the wells breaks the rotational invariance around the NW axis and lifts the degeneracy of those levels.
Clearly, this behavior is exclusively determined by the periodic boundary conditions and the geometric symmetry of the system.
The issues described above for a hexagonal wire can be easily extended to square or triangular NWs.

We now consider the effects of a magnetic field.
In this case the eigenvalue problem is more complicated, but a proper choice of the gauge\cite{Ferrari08} provides an analytical expression of the 2D effective-mass \Sch\ equation, where the dynamics on the surface is still decoupled from the transverse one ({\em i.e.} normal to the curved surface).
We first consider the magnetic levels of a semiconductor shell wrapped around a prismatic NW in an axial magnetic field of intensity $B_{\|}$.
As in the zero field case, the longitudinal and azimuthal directions are uncoupled, and the axial energy dispersion is parabolic.
The $k_y=0$ energy levels, $E_n$, are the eigenvalues of the effective 1D Hamiltonian\cite{Ferrari08} for a charge $q$
\begin{equation}\label{eq:Bpar}
H_{B_{\|}}=\frac{1}{2m^*}\left[-{\hbar}^2 \partial^2_{x}+\im q \hbar B_{\|} d(x) \cos\gamma(x) \partial_{x}+\left(\frac{1}{2}q d(x) \cos\gamma(x) B_{\|}\right)^2\right]+V_{eff}(x),
\end{equation}
which describes the quantum dynamics in the azimuthal direction.
Here, ${d(x)}$ is the distance of $x$ from the center of the polygon, $\gamma(x)$ is the angle defined in \Fig{fig:esawire}.
The familiar cylindrical case $d(x) \cos\gamma(x) = R$ is recovered when $l$ spans one half of the polygon side.\cite{Ajiki93,Meyer07,Ferrari08B}

\begin{figure}
\centering
\includegraphics[width=8cm, angle=0]{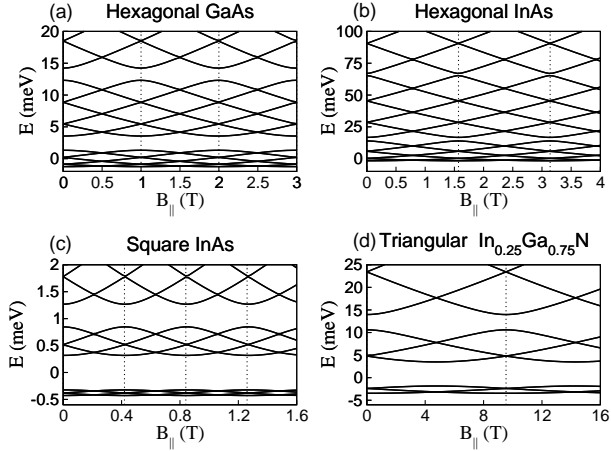}
\caption{\label{fig:axialdens} Magnetic levels $E_{n}$ vs axial field intensity, $B_{\|}$, at $k_y=0$
for a set of core multi-shell NWs.
The values of $B_{\|}$ corresponding to multiples of $\Phi_0$ are indicated by vertical dashed lines.
(a) Hexagonal GaAs shell with $R=40$ nm and $l=2$ nm;\cite{Morral08}
(b) Hexagonal InAs shell with $R=32$ nm and $l=5$ nm;\cite{Tsumura07}
(c) Square InAs shell (see \Reff{Seifert04} for the InP core) with $R=70$ nm and $l=5$ nm;
(d) Triangular In$_{0.25}$Ga$_{0.75}$N shell with $R=18.5$ nm and $l=4$ nm.\cite{Qian04}
}
\end{figure}

From \eqz{eq:Bpar} we have calculated the energy levels $E_n$ as a function of the field intensity for a number of materials and geometries taken from the literature.
In \Fig{fig:axialdens}(a) we show the magnetic levels in the GaAs shell overgrown on the surface of an hexagonal NW with $R=40$ nm and $l=2$ nm (see \Reff{Morral08}).
Note that with an axial field only discrete rotations around the axis are symmetry operations,\cite{Tinkham03} and the magnetic symmetry group is lowered to $C_{6v}$. Accordingly, all energy levels are non degenerate.
Due to the 6-fold symmetry of the system, however, levels form a series of \emph{braids} containing six levels, each one separated from the higher one by an energy gap that depends on $l$, as shown in \Fig{fig:binding}. In \Fig{fig:axialdens}(a) dashed lines indicate the values of the field generating a flux $\Phi$ multiple of the flux quantum $\Phi_0$. The energy levels show clear oscillations with a period $\Phi/\Phi_0$, a periodicity that is typical of AB oscillations, as shown in cylindrical systems.\cite{Olariu85}

In \Fig{fig:axialdens}(b) we show the magnetic levels in the InAs shell\cite{nota-effm} with hexagonal symmetry\cite{Tsumura07} with $R=32$ nm and $l=5$ nm, showing AB oscillations similar to the previous GaAs case. However, here the separation between the braids is smaller since $l/R$ is larger (see the SEM images in \Reff{Tsumura07}), which enhances the tunneling between different edges of the structure.
To highlight the effect of the different symmetry, (\Fig{fig:axialdens}(c)) we calculated the magnetic levels for a hypothetical system, where a low gap layer of InAs is overgrown on the surface of a InP core NW with square symmetry.\cite{Seifert04}
The parameters are $R=70$ nm and $l=5$ nm (see SEM images in \Reff{Seifert04}).
AB oscillations are still present, but, according to the square symmetry of the system, each braid is formed by four levels.

Finally, in \Fig{fig:axialdens}(d), we show the magnetic levels of an In$_{0.25}$Ga$_{0.75}$N shell\cite{nota-effm} with a triangular symmetry, with $R=18.5$ nm and $l=4$ nm.\cite{Qian04} As expected, AB oscillations appear and each braid contains three levels.
We underline the different energy scale and field ranges in the four cases, which depend on the different values of $R$ (influencing the energies $E_{n}$ of the azimuthal modes and the magnetic flux $\Phi$), and the value of $m^*$ that rescales the energies.

We next analyze the effect of a homogeneous magnetic field which is perpendicular to the NW axis with intensity $B_{\bot}$.
As in the previous case, it is still possible to write an effective 1D Hamiltonian\cite{Ferrari08} for the azimuthal dynamics, but now it also depends on the wave-vector $k_y$ along the axis,
\begin{equation}\label{eq:perph}
H_{B\bot}=\frac{1}{2m^*}\left[-{\hbar}^2 \partial^2_{x}+\left(\hbar k_y - q  B_{\bot} s(x)\right)^2\right]+V_{eff}(x).
\end{equation}
Here, $s(x)$ is the distance defined in \Fig{fig:esawire}.
Given the polygonal cross-section, the results will be also dependent on the orientation $\theta$ of the field with respect to the facets.

Since the component of $B_{\bot}$ perpendicular to the surface varies from facet to facet (depending on $\theta$), there is no single magnetic length scale characterizing the system, contrary to the usual planar 2D electron gas case.
As a consequence, the carrier quantum dynamics exhibits two different regimes, that can be analyzed starting from \eqz{eq:perph}.
The Hamiltonian contains a linear term in $B_{\bot} k_y$, and a term proportional to $B^2_{\bot}$.
The former dominates at low field or large longitudinal wave-vector and induces orbital Zeeman splitting.\cite{Ferrari08B} It represents the classical Lorentz force, which induces carrier localization in a quasi-1D channel on one side of the NW, where the vertical component of the field tends to vanish.\cite{nota-side}
For sufficiently high fields, the dominating quadratic term $B^2_{\bot}$ induces carriers localization in two strips located at opposite sides of the NW, where the field is perpendicular to the surface, these states being reminiscent of Landau levels.\cite{nota_cilindri}
These localization mechanisms compete with, or add to the localization along the edges of the NW, depending on the field orientation.

\begin{figure}
\centering
\includegraphics[width=8cm, angle=0]{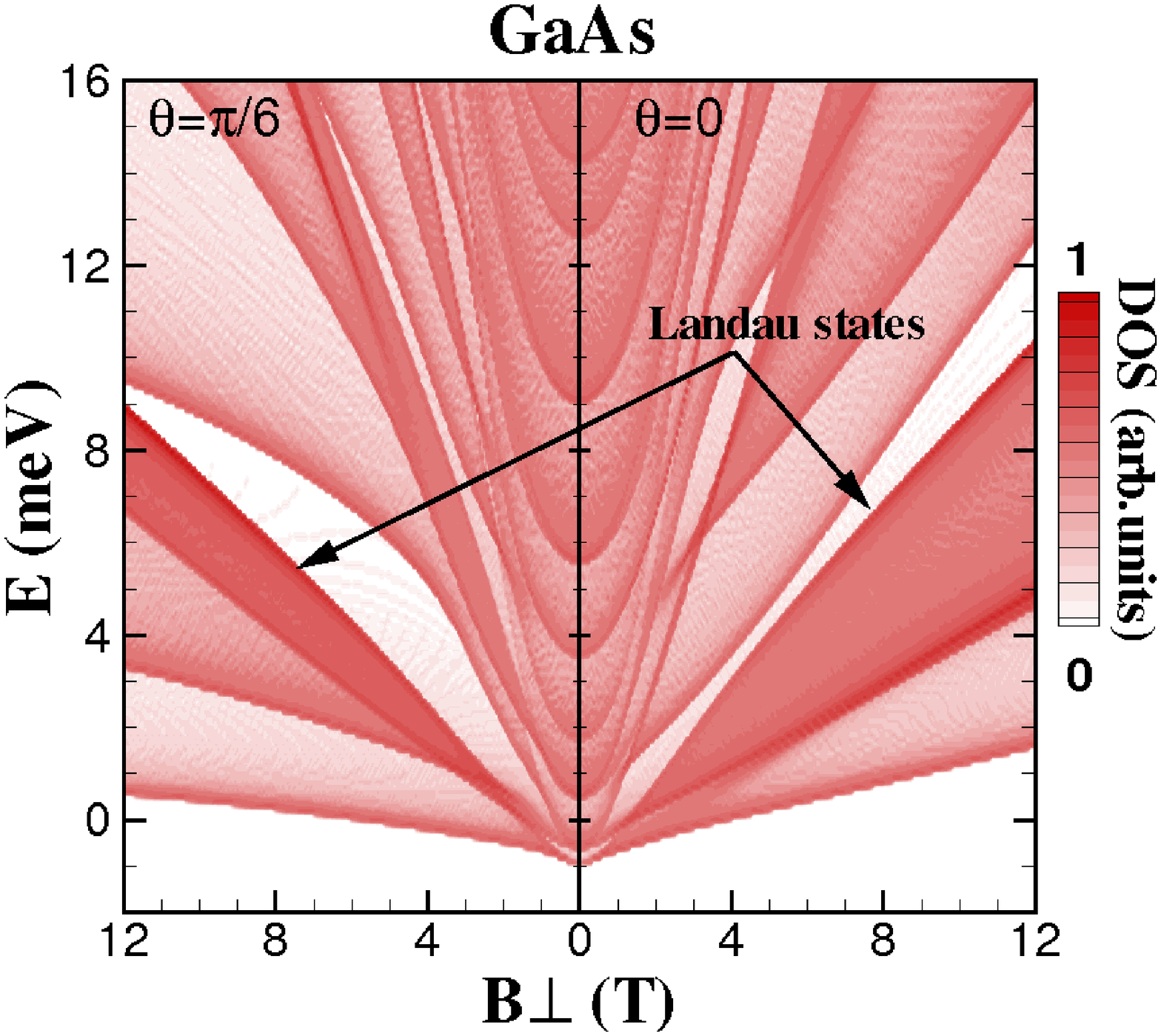}\\
\includegraphics[width=8cm, angle=0]{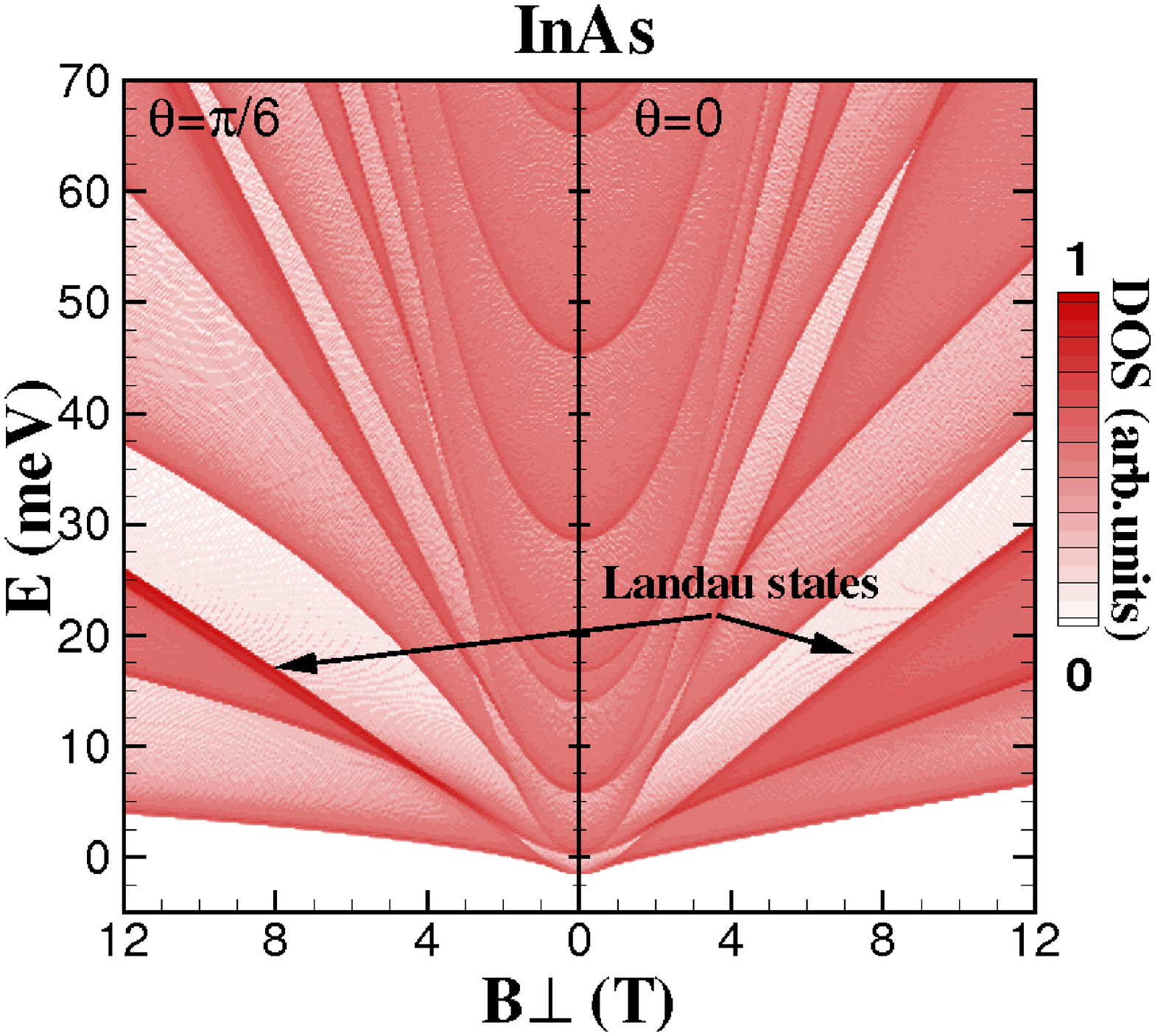}
\caption{\label{fig:perpdens} DOS (gray scale) vs energy, as a function of the intensity of the perpendicular magnetic field $B_\bot$.
Top: hexagonal GaAS with $R=40$ nm and $l=2$ nm;
Bottom: hexagonal InAs with $R=32$ nm and $l=5$ nm.
Left panels: field oriented at $\theta = \pi/6$. Right panels: field oriented at $\theta =0$.
(see \Fig{fig:esawire} for the definition of $\theta$).
}
\end{figure}

By solving \eqz{eq:perph}, we have computed the density of states (DOS) shown in \Fig{fig:perpdens}, for the GaAs (top) and InAs (bottom) shell with hexagonal symmetry. Similar results can be easily inferred for the square or triangular systems discussed above.
For each system, the DOS has been calculated with two limiting orientations of the magnetic field: $\theta = 0$ and $\theta = \pi/6$ (see \Fig{fig:esawire}).
At each magnetic field, the magnetic energy levels form a set of subbands with a minimum (darker regions in \Fig{fig:perpdens})  which, in general, corresponds to a state with $k_y \ne 0$, due to the linear term in $B_{\bot} k_y$.
The lowest subbands (below the levels labeled as Landau states) correspond to states which tend to be localized on one side of the NW with respect to the field direction, that is where the orthogonal component of $B_\bot$ on the surface of the NW vanishes.
For $\theta=0$ (right panels) the localization area coincides with an edge of the NW. Therefore, edge-induced localization is enhanced by the field in one of the side edges of the NW, while the other edges are charge depleted. For $\theta=\pi/6$ (left panels), on the other hand, the area of field-induced localization coincides with one of the NW facets, and all edges are charge depleted. Note the differences between the GaAs and InAs results, which are due to the different tunnel-coupling between the edge wells.
At higher energies, peaks labeled as Landau states shift linearly with $B_{\bot}$. These states are confined by the field on the top and the bottom of the NW with respect to the field direction. Again, if these regions correspond to edges of the NW (that is for $\theta=\pi/6$), localization is enhanced for  these two edges, while the other four edges are charge depleted. On the other hand, if $\theta=0$ carriers tend to be confined in the center of the two facets where the field is perpendicular, and all edges of the NW are charge depleted.
Note the different slopes of the energy levels with field intensity for different orientation $\theta$ of the field. In fact, only the orthogonal component $\cos(\theta) B_{\bot}$ is effective, and the Landau-level energy is correspondingly smaller. 
For higher energies, the succession of peaks is repeated. A discussion of these states in a related cylindrical systems is given in \Reff{Ferrari08B}. 

In conclusion, we have shown that in core multi-shell NWs quantum confinement occurs in a set of quasi-1D channels located at the edges of the NW, and we have evaluated such confinement for typical parameters.
An axial magnetic field induces AB oscillations of the multiplets arising from the prismatic symmetry. A transverse magnetic field may enhance edge localization or deplete these channels and induce Landau level formation along the facets or edges of the NW, depending on the field direction with respect to the NW facets.

This work has been partially supported by project FIRB-RBIN04EY74 and Cineca Calcolo Parallelo 2008.

\bibliography{ferrari}

\begin{thebibliography}{42}
\bibitem[{not({\natexlab{a}})}]{nota-bent}
\bibinfo{note}{This is also evident if one thinks of a bent as a localizing 1D
  defect in an otherwise translational invariant 2D system.}
\bibitem[{not({\natexlab{b}})}]{nota-effm}
\bibinfo{note}{We model conduction electrons using $m^* = 0.067m_e$ for GaAs,
  $m^* = 0.023m_e$ for InAs, $m^* = 0.125m_e$ for In$_{0.25}$Ga$_{0.75}$N,
  where $m_e$ is the free electron mass. Results for holes differ only for a
  scaling factor in the energy plots.}
\bibitem[{not({\natexlab{c}})}]{nota-side}
\bibinfo{note}{Which side is determined by the sign of $q$, the direction of
  $k_y$ and the orientation of $B_{\bot}$.}
\bibitem[{not({\natexlab{d}})}]{nota_cilindri}
\bibinfo{note}{The same magnetic effects have been described in cylindrical
  systems, see \Reff{Ajiki93,Ferrari08B}.}
\end{thebibliography}

\end{document}